\documentstyle[pre, aps, epsfig]{revtex}
\begin{document}
\maketitle
\begin{flushright} {\bf \Large I30} \end{flushright}
\vspace{0.8cm}

\begin{center}
{\bf\Large Surface effects in nanoparticles} \\
\vspace{0.8cm}

H.\ Kachkachi$^a \footnote{{\it Corresponding author~: kachkach@physique.uvsq.fr}}
$, M.\ Nogu\`{e}s$^a$, E.\ Tronc$^b$, and D.A.\ Garanin$^c$\\
\vspace{0.15cm}

%\small
$^a$LMOV-CNRS UMR 8634, Universit\'{e} de Versailles St. Quentin,\\
45, avenue des Etats-Unis, 78035 Versailles Cedex, France \\
%\vspace{0.15cm}
$^b$LCMC-CNRS URA\ 1466, Universit\'{e} Pierre \& Marie Curie,\\
4 place Jussieu, 75252 Paris Cedex, France\\ 
%\vspace{0.15cm}
$^c$Max-Planck-Institut f\"ur Physik komplexer Systeme,\\
N\"othnitzer Strasse 38, D-01187 Dresden, Germany.
\end{center}
\maketitle
\begin{abstract}
We study the finite-size and surface effects on the
thermal and spatial behaviors of the magnetisation of a small
magnetic particle. We consider two systems:
1) A box-shaped particle of simple cubic structure with either periodic or
free boundary conditions. This case is treated analytically using the
isotropic model of $D$-component spin vectors in the limit $D\rightarrow
\infty ,$ including the magnetic field.
2) A more realistic particle ($\gamma $-Fe$_{2}$O$_{3}$) of ellipsoidal (or
spherical) shape with open boundaries. The magnetic
state in this particle is described by the anisotropic classical
Dirac-Heisenberg model including exchange and dipolar interactions, and bulk
and surface anisotropy. This case is dealt with by the classical Monte Carlo
technique.
%{\bf PACS~: 75.30.Pd} (surface magnetism){\bf , 75.50.Gg} (ferrimagnetism)%
%{\bf , 75.60.Jp} (nanoparticles).
\vspace{0.2cm}
\end{abstract}
\section{Introduction}
From the physical point of view, nanoparticles exhibit such interesting
features as superparamagnetism and exponentially slow relaxation rates 
at low temperatures due to anisotropy barriers. However, the picture
of a single-domain magnetic particle where all spins are pointing into the
same direction, leading to coherent relaxation processes, ceases to be 
valid for very small particles where surface effects become really crucial. 
For instance, in a particle of radius $\sim 4$ nm, $50\%$ of atoms lie 
on the surface.
Therefore, it is necessary to understand 
the effect of free boundaries first on the static and then on the 
dynamical properties of nanoparticles.
However, one of the difficulties which is inherent to systems of round 
(spherical or ellipsoidal) geometries, consists in separating surface
effects due to symmetry breaking of the crystal field on the boundaries 
and the unavoidable finite-size effects caused by using systems of 
finite size. 
In hypercubic systems, this problem is easily handled by using periodic 
boundary conditions, but this is not possible in other topology, and 
thus surface and finite-size effects are mixed together.

In this article, we discuss surface and finite-size effects on the 
thermal and spatial behaviours of the intrinsic magnetisation of an 
isolated small particle. We consider two different systems: 
1) A cube of simple cubic structure with either
periodic or free boundary conditions. This system is treated 
analytically by the isotropic model of $D$-component spin vectors in 
the limit $D\rightarrow
\infty ,$ in magnetic field\cite{HKDG}. 2) The second system, which is 
more realistic, is the maghemite particle ($\gamma $-Fe$_{2}$O$_{3}$) 
of ellipsoidal (or spherical) shape with open boundaries. 
The appropriate model is the anisotropic classical Dirac-Heisenberg
model including exchange and dipolar interactions, and taking account of
bulk and surface anisotropy. On the contrary, this system can only be 
dealt with using numerical approaches such as the classical Monte Carlo 
technique
\cite{HK}.
In the case of a cubic system we obtain the thermal behaviour of local
magnetisations at the center of faces, edges and corners. 
An exact and very useful relation between the intrinsic magnetisation 
and the magnetisation induced by the magnetic field, valid at all 
temperatures and fields, was obtained in Ref.\cite{HKDG}. It was shown 
that the positive contribution of finite-size effects to the
magnetisation is lower than the negative one rendered by boundary effects, 
thus leading to a net decrease of the magnetisation with respect to the bulk.
For the maghemite, this study has been performed in a very small and
constant magnetic field; the surface shell is assumed to be of constant
thickness and only the particle size is varied. So, the thermal 
behaviour of the intrinsic magnetisation is obtained for different
particle sizes \cite{HK}. This behaviour is compared with that of a 
cubic maghemite particle with periodic boundary conditions but without 
anisotropy. In this case the contributions of finite-size and surface 
effects lead to the same results as for the cube system, but the 
difference between them is now much larger, due to surface anisotropy. 
In addition, we show that the magnetisation profile is temperature 
dependent.
\section{Cubic system: $D\rightarrow \infty $ spherical model}
We consider an isotropic box-shaped magnetic system of volume ${\cal N}%
=L^{3}$, with simple-cubic lattice structure, and
nearest-neighbour exchange coupling, in a uniform magnetic field. For this we 
use the Hamiltonian of the isotropic classical $D$-component
vector model\cite{Garanin}, that is, 
\begin{equation}
{\cal H}=-{\bf h\cdot }%
\mathop{\displaystyle \sum }%
\limits_{i}{\bf s}_{i}-\frac{1}{2}%
\mathop{\displaystyle \sum }%
\limits_{i,j}\lambda _{ij}%
\mathop{\displaystyle \sum }%
\limits_{\alpha =1}^{D}s_{\alpha i}s_{\alpha j},  \label{DHam}
\end{equation}
where ${\bf s}_{i}$ is the normalized $D$-component vector, $\left| {\bf s}%
_{i}\right| =1$; ${\bf h\equiv H/}J_{0}$ is the magnetic field, and $\lambda
_{ij}\equiv J_{ij}/J_{0}$ the exchange coupling. We also define the reduced
temperature $\theta \equiv T/T_{c}^{MFA},$ $T_{c}^{MFA}=$ $J_{0}/D$ being
the Curie temperature of this model in the mean-field approximation, $J_{0}$
is the zero-momentum Fourier component of $J_{ij}$. In this model, the
magnetisation ${\bf m}$ is directed along the field ${\bf h,}$ so that ${\bf %
h=}h{\bf e}_{z}$ and ${\bf m}_{i}{\bf =}m_{i}{\bf e}_{z}.$ Using the diagram
technique for classical spin systems \cite{Garanin} in the limit $%
D\rightarrow \infty ,$ generalizing it so as to include the magnetic field
and adopting a matrix formalism\cite{HKDG}, one ends up with a closed system
of equations for the average magnetisation component $%
m_{i}\equiv \left\langle s_{zi}\right\rangle $ and correlation functions $%
s_{ij}\equiv D\left\langle s_{\alpha i}s_{\alpha j}\right\rangle $ 
with $\alpha \geq 2$\cite{HKDG}, 
\begin{equation}
\sum_{j}{\cal D}_{ij}m_{j}=G_{i}h,\qquad \sum_{j}{\cal D}_{ij}s_{jl}=\theta
G_{i}\delta _{il},  \label{DefMatr}
\end{equation}
where ${\cal D}_{ij}\equiv \delta _{ij}-G_{i}\lambda _{ij}$ is the Dyson
matrix of the problem, and $G_{i}$ is a local function to be determined from
the set of constraint equations on all sites $i=1,\ldots ,{\cal N}$ of the
lattice 
\begin{equation}
s_{ii}+{\bf m}_{i}^{2}=1.  \label{Constraint}
\end{equation}

Now, we define the induced average magnetisation per site by 
\begin{equation}
{\bf m=}\frac{1}{{\cal N}}%
%TCIMACRO{\dsum}
%BeginExpansion
\mathop{\displaystyle \sum }%
%EndExpansion
\limits_{i}{\bf m}_{i}  \label{m}
\end{equation}
which vanishes for finite-size systems in the absence of magnetic field due
to the Golstone mode associated with global rotations of the
magnetisation. On the other hand, it is clear that at temperatures $\theta
\ll 1$ the spins in the system are aligned with respect to each other and
there should exist an {\it intrinsic} magnetisation. The latter is usually
defined for finite-size systems as 
\begin{equation}
M=\sqrt{\left\langle \left( \frac{1}{{\cal N}}%
%TCIMACRO{\dsum}
%BeginExpansion
\mathop{\displaystyle \sum }%
%EndExpansion
\limits_{i}{\bf s}_{i}\right) ^{2}\right\rangle }=\sqrt{{\bf m}^{2}+\frac{1}{%
{\cal N}^{2}}%
%TCIMACRO{\dsum}
%BeginExpansion
\mathop{\displaystyle \sum }%
%EndExpansion
\limits_{i,j=1}^{{\cal N}}s_{ij}},  \label{M}
\end{equation}
where the second equality is valid in the limit $D\rightarrow \infty .$
Note that $M\geq m$ and that $M$ remains non zero for $h=0$; in this case in
the limit $\theta \rightarrow 0,$ $s_{ij}=1$ for all $i$ and $j,$ and $%
M\rightarrow 1.$ For $\theta \rightarrow \infty $ the spins become
uncorrelated and $M\rightarrow 1/\sqrt{{\cal N}}.$
In the limit of ${\cal N}\rightarrow \infty ,$ the intrinsic magnetisation $M
$ approaches that of the bulk system.
In the presence of a magnetic field, the Goldstone mode is suppressed and
the magnetisation ${\bf m}$ of Eq.(\ref{m}) no longer vanishes, this is 
why we call it the {\it supermagnetisation,} in contrast
with the intrinsic magnetisation $M$. If the field is strong the magnitude
of the supermagnetisation approaches the intrinsic magnetisation.

An important exact relation was established in Ref.\cite{HKDG} between 
$M$ and $m,$
\begin{equation}
m=M\frac{2{\cal N}Mh/\theta }{1+\sqrt{1+(2{\cal N}Mh/\theta )^{2}}}=MB({\cal %
N}MH/T),  \label{mvsM}
\end{equation}
where $B(\xi )=(2\xi /D)/\left[ 1+\sqrt{1+(2\xi /D)^{2}}\right] $ is the
Langevin function for $D\gg 1$. Note that Eq.\ (\ref{mvsM}) is usually
applied to superparamagnetic systems with the spontaneous bulk magnetisation 
$m_{{\rm b}}(T)$ in place of $M(T,H)$. However, unlike $m_{{\rm b}}(T)$, the
intrinsic magnetisation $M$ of Eq.\ (\ref{M}) is a {\em pertinent}
characteristic of a finite magnetic system and depends on both field and
temperature.

Solving the model above consists in determining ${\bf m}_{i}$ and $s_{ij}$
as functions of $G_{i}$ from the linear equations (\ref{DefMatr}), and
inserting these solutions in the constraint equation (\ref{Constraint}) 
in order to obtain $G_{i}$. 
Two types of boundary conditions are considered, free boundary 
conditions (fbc) and periodic boundary conditions (pbc). 
In the case of fbc, ${\bf m}_{i}$ and $G_{i}$ are inhomogeneous and $%
s_{ij}$ non-trivially depends on both indices due to boundary effects. In
this case the exact solution is found numerically, though some analytic
calculations can be performed at low temperature and field.
Whereas in the pbc case the solution becomes homogeneous and the problem
greatly simplifies. 
Although the model with pbc is unphysical, it allows for an
analytical treatment and study of finite-size effects separately
from boundary effects.

At low temperature, the intrinsic magnetisation in the 
fbc case, including only the contributions from faces, reads\cite{HKDG},
\begin{equation}
M\cong 1-\frac{\theta W}{2}\left[ 1-\Delta _{{\cal N}}+\frac{6}{5}
\frac{1}{L} \right] ,
\label{MLTRes}
\end{equation}
where $W$ is the well-known Watson's integral and 
\begin{equation}
\Delta _{{\cal N}}=\frac{1}{W}\left( W-\frac{1}{{\cal N}}
\sum_{{\bf q\neq 0}%
}{}\frac{1}{1-\lambda _{{\bf q}}}\right) >0 
\end{equation}
describes the finite-size effects, with $\Delta _{{\cal N}} \propto 1/L$, while the last term in (\ref{MLTRes})
represents the contribution from boundaries. The first term, on the other hand, 
is the bulk contribution which survives in the limit $L\to\infty$.
In contrast with the
finite-size effects, boundary effects entail a decrease of the intrinsic
magnetisation. The contributions to Eq.\ (\ref{MLTRes}) from the edges and
corner are of order $\theta /L^{2}$ and $\theta
/L^{3}$, respectively.
%
%%%%%%%%%%%%%%%%%%%%%%%%%%%%%%%%%%%%%%%%%%%%%%%%%%%%%%%%%%%%%%%%%%%%%%%%%%%%%%%%%%%%%%%%%%%%%%
\begin{figure}[h]
\begin{center}
\epsfig{file=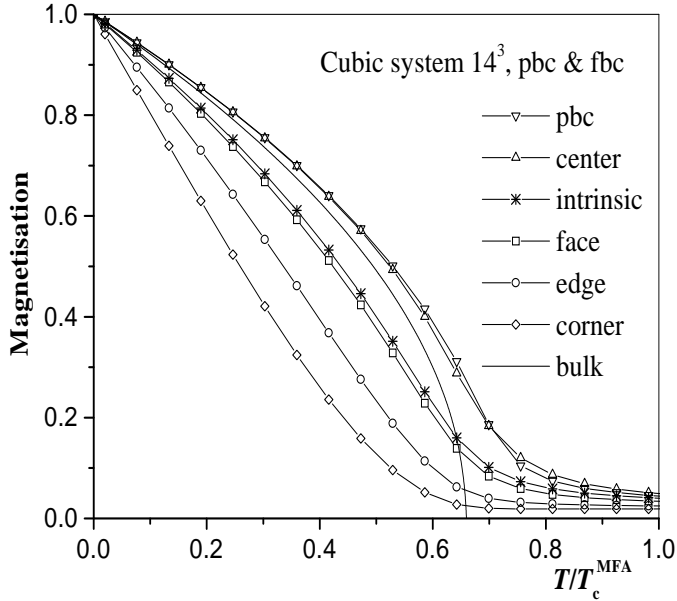, width=15cm,height=12cm}
\end{center}
\vspace{-3.8cm}
\caption{\label{Barce1}
Temperature dependence of the intrinsic magnetisation $M$, Eq.\ (\protect\ref{M}),
and local magnetisations of the $14^3$ cubic system with free
and periodic boundary conditions in zero field.}
\end{figure}
%%%%%%%%%%%%%%%%%%%%%%%%%%%%%%%%%%%%%%%%%%%%%%%%%%%%%%%%%%%%%%%%%%%%%%%%%%%%%%%%%%%%%%%%%%%%%
%
Fig.\ \ref{Barce1} shows the temperature dependence of the intrinsic
magnetisation $M$ , Eq.\ (\ref{M}), and local magnetisations of the $14^{3}$
cubic system with free and periodic boundary conditions in zero field. For
periodic boundary conditions, $M$ exceeds the bulk magnetisation at all
temperatures. In particular, at low temperatures this agrees with the
positive sign of the finite-size correction to the magnetisation, Eq.\ (\ref
{MLTRes}). The magnetisation at the center of the cube with free boundary
conditions is rather close to that for the model with pbc in the whole
temperature range and converges with the latter at low temperatures. Local
magnetisations at the center of the faces and edges and those at the corners
decrease with temperature much faster than the magnetisation at the center.
This is also true for the intrinsic magnetisation $M$ which is the average
of the local magnetisation $M_{i}$ over the volume of the system. 
One can see that, in the
temperature range below the bulk critical temperature, $M$ is smaller than
the bulk magnetisation. This means that the boundary effects suppressing $M$
are stronger than the finite-size effects which lead to the increase of the
latter, and this is in agreement with the low-temperature formula of Eq.\ (\ref
{MLTRes}).
\section{Maghemite particles: Monte Carlo simulations}
In this section, we consider the more realistic case of (ferrimagnetic)
maghemite nanoparticles ($\gamma $-Fe$_{2}$O$_{3}$) of ellipsoidal (or
spherical) shape with open boundaries, in a very
small and uniform magnetic field. The surface shell is assumed to be
of constant thickness ($\sim 0.35$ nm) 
in\cite{Tronc}, and only the particle size is varied\cite{HK}.

To deal with
spatial magnetisation distributions\cite{Akhiezer} one has to consider 
exchange, anisotropy and magneto-static energies together. Accordingly, 
our model for a nanoparticle is the classical Dirac-Heisenberg 
Hamiltonian including exchange and dipole-dipole interactions, 
anisotropy, and Zeeman contributions. 
Denoting (without writing explicitly) the dipole-dipole
interaction by $H_{dip}$, our model reads, 
\begin{eqnarray}
{\cal H} &=&-%
\mathop{\displaystyle \sum }%
\limits_{i,{\bf n}}%
J_{\alpha \beta }
\mathop{\displaystyle \sum }%
\limits_{\alpha ,\beta }{\bf S}_{i}^{\alpha }\cdot {\bf S}%
_{i+{\bf n}}^{\beta }-K%
\mathop{\displaystyle \sum }%
\limits_{i=1}^{N_{t}}\left( {\bf S}_{i}\cdot {\bf e}_{i}\right) ^{2}
\label{DH}-(g\mu _{B})H%
\mathop{\displaystyle \sum }%
\limits_{i=1}^{N_{t}}{\bf S}_{i}+H_{dip}  \nonumber
\end{eqnarray}
where $J_{\alpha \beta }$ are the exchange couplings between
nearest neighbours spanned by the unit vector ${\bf n}$; ${\bf S}_{i}^{\alpha
}$ is the (classical) spin vector of the $\alpha ^{th}$ atom at site $i;$ $H$
is the uniform field applied to all spins in the
particle, $K>0$ is the anisotropy constant and ${\bf e}_{i}$ the single-site
anisotropy axis. In both cases of a spherical and ellipsoidal particle, we
consider a uniaxial anisotropy in the core along our $z$ reference axis
(major axis for the ellipsoid), and single-site anisotropy on the
surface, with equal anisotropy constant $K_{s},$ and ${\bf e}_{i}$
are defined so as to point outward and normal to the surface \cite{Kodama}.

Our method of simulation proceeds as follows: we start with a regular box 
of spinel structure, then we cut in a sphere or an ellipsoid that
contains
the total number $N_{t}$ of spins of a given particle. We distinguish between spins 
in the core (of number $%
N_{c}$) from those on the surface ($N_{s}$) of the particle according to
whether or not their coordination number is equal to  that of a system with periodic
boundary conditions (pbc). All
spins in the core and on the surface are identical but interact via
different couplings; exchange interactions between the core and surface
spins are taken equal to those inside the core. Our parameters are as
follows: the exchange interactions are (in units of K) $%
J_{AB}/k_{B}\simeq -28.1,J_{BB}/k_{B}\simeq -8.6,J_{AA}/k_{B}\simeq -21.0$.
The bulk and surface anisotropies are $k_{c}\equiv (K_{c}/k_{B})\simeq
8.13\times 10^{-3},$ $k_{s}\equiv (K_{s}/k_{B})\simeq 0.5$, respectively,
where $k_{B}$ is the Boltzmann constant.

\begin{figure}[h]
\begin{center}
\epsfig{file=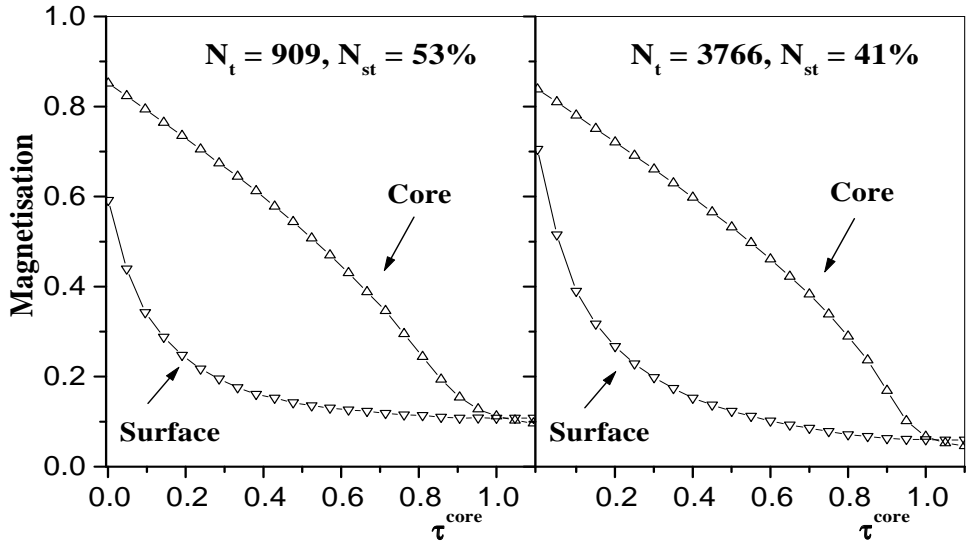, width=19cm,height=12cm}
\end{center}
\vspace{-4.5cm}
\caption{\label{Barce2}
Temperature dependence of the surface and core magnetisations for 
$N_t=909$ and $3766$.}
\end{figure}

In Fig.\ \ref{Barce2}, we plot the thermal variation of the core and 
surface
contributions to the magnetisation (per site) as a function of the 
reduced temperature $\tau ^{core}\equiv T/T_{c}^{core}$ for 
$N_{t}=909,3766$ corresponding to $N_{st}\equiv N_s/N_t=53\%,41\%$
and diameters of 
circa $4$ and $6$ nm, respectively. The core and surface magnetisations are
averages over all spins in the core or on the surface, respectively.
For both sizes we see that the surface
magnetisation $M_{surf}$ decreases more rapidly than the core contribution $%
M_{c}$ as the temperature increases, and has a positive curvature while that
of $M_{c}$ is negative. Moreover, it is seen that even the (normalised) core
magnetisation per site does not reach its saturation value of $1$ at very
low temperatures, which shows that the magnetic order in the core is 
disturbed by the surface (see Fig.4 below). As the size decreases the maximum 
value of $M_{surf}$ decreases showing that the magnetic disorder is enhanced.

\begin{figure}[h]
\begin{center}
\epsfig{file=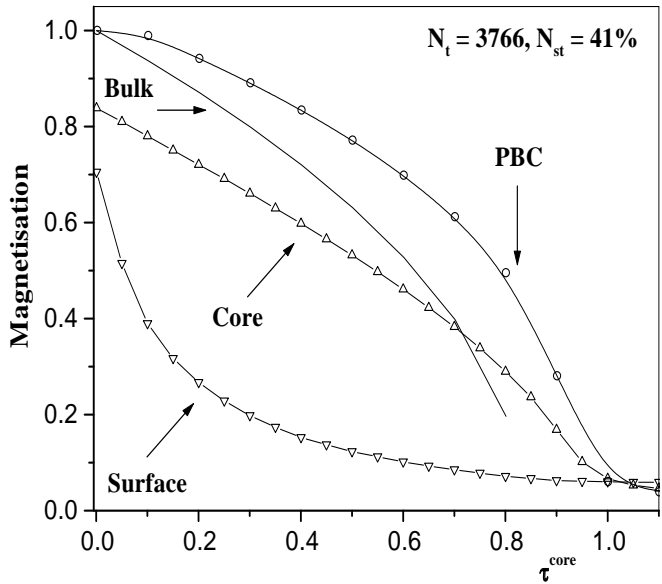, width=15cm,height=12cm}
\end{center}
\vspace{-4cm}
\caption{\label{Barce3}
Temperature dependence of the surface and core magnetisations for $N_t=3766$, 
magnetisation of the bulk system ($N_t=\infty$), and the magnetisation of the cube with the 
spinel structure and periodic boundary conditions (PBC) with $N_t=40^3$.}
\end{figure}

In Fig.\ \ref{Barce3} we plot the core and surface magnetisations 
(with $N_t = 909, 3766$, and $N_{st}=53 \%, 41\%$), 
the magnetisation of a cube with spinel
crystalline structure and pbc, and the bulk
magnetisation as functions of the reduced temperature $\tau^{core}$. 
Apart from the obvious shift to lower temperatures of the
critical region due to the finite-size and surface effects, we see that, as
was also shown analytically for the cube system, the finite-size
effects give a positive contribution to the magnetisation with respect to
the bulk, whereas the surface effects yield a negative contribution.
Moreover, it is seen that for nanoparticles the contribution from the
surface is much larger than that coming from finite-size effects. The
difference between the two contributions appears to be enhanced by the
surface anisotropy in the case of nanoparticles.

\begin{figure}[h]
\begin{center}
\epsfig{file=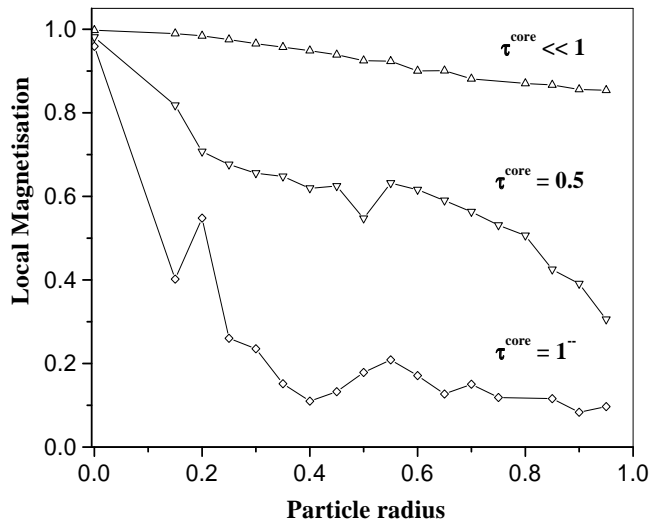, width=10cm,height=15cm}
\end{center}
\vspace{-7.5cm}
\caption{\label{Barce4}
Spatial variation of the net magnetisation of a spherical nanoparticle of $3140$ spins,
as a function of the normalised particle radius, for $\tau^{core}\ll 1$, and 
$\tau^{core} = 0.5$, $\tau^{core} \simeq 1^-$.}
\end{figure}

In Fig.\ \ref{Barce4} we plot the spatial evolution of the orientation 
of the magnetic moment from the center to the border of the particle, 
at different temperatures. 
At all temperatures, the magnetisation decreases with
increasing particle radius. This obviously suggests that the magnetic
disorder starts at the surface and gradually propagates into the core of the
particle.
At high temperatures, the local
magnetisation exhibits a jump of temperature-dependent height, and 
continues to decrease. This indicates that there is a temperature-dependent radius, 
smaller than the particle radius, within which the magntisation assumes relatively
high values. This result agrees with that of Ref.\cite{Wildpaner} (for spherical
nanoparticles with simple cubic structure) where this radius 
was called the {\it magnetic radius}.
The local magnetisation also depends on the direction of the radius vector,  
especially in an ellipsoidal particle.
%
%%%%%%%%%%%%%%%%%%%%%%%%%%%%%%%%%%%%%%%%%%%%%%%%%%%%%%%%%%%%%%%%%%%%%%%%%%%%%%%%%%%%%%%%%%%%%%
%
%%%%%%%%%%%%%%%%%%%%%%%%%%%%%%%%%%%%%%%%%%%%%%%%%%%%%%%%%%%%%%%%%%%%%%%%%%%%%%%%%%%%%%%%%%%%%
%
%%%%%%%%%%%%%%%%%%%%%%%%%%%%%%%%%%%%%%%%%%%%%%%%%%%%%%%%%%%%%%%%%%%%%%%%%%%%%%%%%%%%%%%%%%%%%%
%%%%%%%%%%%%%%%%%%%%%%%%%%%%%%%%%%%%%%%%%%%%%%%%%%%%%%%%%%%%%%%%%%%%%%%%%%%%%%%%%%%%%%%%%%%%%
%
%%%%%%%%%%%%%%%%%%%%%%%%%%%%%%%%%%%%%%%%%%%%%%%%%%%%%%%%%%%%%%%%%%%%%%%%%%%%%%%%%%%%%%%%%%%%%%
%%%%%%%%%%%%%%%%%%%%%%%%%%%%%%%%%%%%%%%%%%%%%%%%%%%%%%%%%%%%%%%%%%%%%%%%%%%%%%%%%%%%%%%%%%%%%
%
\section{Conclusion}
Both for the cube system and the nanoparticle of the maghemite type 
surface effects yield a negative contribution to the
intrinsic magnetisation, which is larger than the positive contribution 
of
finite-size effects, and this results in a net decrease of the 
magnetisation with
respect to that of the bulk system. In the first case we have been able 
to separate finite-size effects from surface effects by considering the 
same system with periodic and free boundary conditions. On the other 
hand, the results for
a spherical or ellipsoidal nanoparticle with free boundaries have been
compared to those of a cube with a spinel structure and periodic
boundary conditions, but without any anisotropy. In this case, it turns
out that the
contributions from surface and finite-size effects have the same sign as
before but the difference between them becomes larger, due to surface
anisotropy.

These spin models invariably predict that the surface magnetisation (per spin) of systems
with free boundaries is smaller
than the magnetisation of the bulk system. However, experiments on layered
systems, especially of 3d elements, have shown that there is enhancement 
of the magnetic moment on the surface, which has been attributed to the contribution 
of orbital moments\cite{Eriksson}. It is clear that the models presented here do not 
account for such effect, but they can be generalised so as to include orbital as 
well as spin vectors.

\end{document}